\documentclass[aps,prl,twocolumn,groupedaddress,showpacs]{revtex4-1}

\usepackage{amsmath,amssymb}
\usepackage[dvipdfmx]{graphicx}

\newcommand{\bra}[1]{\langle{#1}|}
\newcommand{\ket}[1]{|{#1}\rangle}

\begin{document}

\preprint{UT-13-40}

\title{Bound State Effect on the Electron $g-2$}

\author{Go Mishima}
\email[]{mishima@hep-th.phys.s.u-tokyo.ac.jp}
\affiliation{Department of Physics, The University of Tokyo, Tokyo 113-0033, Japan}

\date{\today}

\begin{abstract}
We evaluate a non-perturbative QED contribution to the electron $g-2$ which comes from virtual positronium.
We find it to be 1.3$(\alpha /\pi )^5\simeq 9.0\times 10^{-14}$.
This value is comparable to the five-loop contribution of usual perturbative calculation,
and several times larger than the electroweak correction.
\end{abstract}

\pacs{11.10.St, 12.20.Ds, 13.40.Em, 14.60.Cd}

\maketitle

\section{Introduction}
\begin{figure}[b]
\begin{center}
\includegraphics[width =2cm, bb=246 437 474 571]{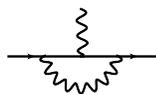}
\caption{Feynman diagram corresponding to the leading contribution to the electron $g-2$.}\label{g-2}
\end{center}
\end{figure}

One of the best agreement between measurement and theory comes from
the anomalous magnetic moment ($g-2$) of electron.
It was first observed in precision spectroscopy of various atoms
\cite{PhysRev.71.914}, \cite{PhysRev.72.971}, \cite{PhysRev.72.1256.2} in 1947,
and first calculated based on quantum electrodynamics (QED) \cite{PhysRev.73.416} in 1948 (FIG. \ref{g-2}).
Since then, the electron $g-2$ has played a central role in testing the validity of QED.
At present, the theoretical prediction based on the standard model consists of three parts --
the dominant QED contribution,
small hadronic correction, and tiny electroweak correction (TABLE 1).
Recently, the five-loop calculation of QED was completed \cite{Aoyama:2012wj}, and
the hadronic correction \cite{Nomura:2012sb}, \cite{Prades:2009tw} and the electroweak correction \cite{Mohr:2012tt}
are also evaluated very precisely.
In fact, the dominant theoretical uncertainty comes from parameter determination of the fine structure constant \cite{Bouchendira:2010es},
and its current relative uncertainty is $6.6\times 10^{-10}$.
On the experimental side, 
the relative uncertainty of the current measured value is $2.4\times 10^{-10}$ \cite{2008PhRvL.100l0801H},
which is several times smaller than that of theory.

Due to its great precision, 
the electron $g-2$ is used to constrain new physics models beyond the standard model
\cite{Jaeckel:2010xx}, \cite{Giudice:2012ms}, \cite{Endo:2012hp}.
These studies rely on the premise that the calculation based on the standard model is already settled.
However, it is worth reconsidering the standard model prediction more delicately
because apparently negligible contribution can compete with current precision,
which is greatly improved in recent years.

In this letter,
we discuss a hitherto neglected QED contribution to the electron $g-2$,
which originates from non-perturbative effect on vacuum polarization.
Vacuum polarization is generated by virtual pair creation of electron and positron, and
if these two particles are bound, then they appear as one stable particle -- positronium.
Virtual positronium contribution never arise in the traditional perturbation expansion from the tree level approximation
and we need some resummation.
We argue that the bound state contribution calculated in this letter is distinct from the perturbative contribution of finite order,
therefore we need to add both contributions.

\section{Calculation}
\begin{figure}[b]
\begin{center}
\includegraphics[width =7cm, bb=72 675 540 720]{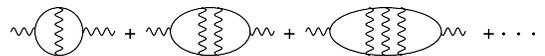}
\caption{Ladder diagrams appearing in the electron loop of the vacuum polarization correction.}\label{ladder}
\end{center}
\end{figure}

Here we consider QED of photons and electrons only.
The vacuum polarization function is defined as
\begin{eqnarray}
\mathrm{i} q^2\Pi (q^2)Q^{\mu\nu}  \equiv 
\int \mathrm{d} ^4x~ \mathrm{e}^{\mathrm{i} qx}
\bra{0} {\rm T} \{j^\mu (x)j^\nu (0) \} \ket{0} _{\rm 1PI} \label{pi}
\end{eqnarray}
where $j^\mu (x)\!\! =\!\! -e\bar\psi (x)\gamma ^\mu \psi (x)$ is the electromagnetic current,
$Q^{\mu\nu}=g^{\mu\nu}-q^\mu q^\nu /q^2$ is the projection matrix,
and the subscript ``1PI" means that the one-particle-irreducible parts are evaluated.
For the definitions and the notations, we follow ref.\cite{Peskin:1995ev}.
In the ordinary perturbation theory, the vacuum polarization function are expanded in terms of the fine structure constant $\alpha \simeq 1/137$,
and these contributions to the electron $g-2$ are studied very precisely \cite{Aoyama:2012wj}, \cite{Laporta:1900zz}, \cite{Baikov:2013ula}.
However, if the electrons in the loop of the ladder diagrams (FIG. \ref{ladder}) are non-relativistic, 
the propagators of those electrons provide an enhancement factor $\alpha ^{-1}$ which cancels the perturbation parameter \cite{Berestetsky:1982aq}.
In this case, we cannot rely on the usual perturbation theory truncated to some fixed finite order.
Instead, we have to resum ladder diagrams \cite{Cornwall:1974vz}, 
and the resummation of ladder diagrams generates the bound state of electron and positron \cite{GellMann:1951rw}, \cite{Salpeter:1951sz}.
This means that there is a mixing between the photon and the positronium.

\begin{figure}[t!]
\begin{center}
\includegraphics[width =5cm, bb=0 0 1187 691]{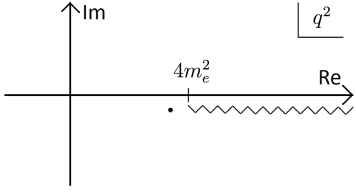}
\caption{Branch cut and singularities of $\Pi (q^2)$ in the complex plane of $q^2$.
The branch cut is present in ordinary perturbation calculation,
and the singularities are generated form the bound state effect.
}\label{plane}
\end{center}
\end{figure}

The vacuum polarization function acquires additional poles at the masses of the positronium states from the non-perturbative effect explained above.
In FIG.~\ref{plane}, the branch cut and the poles of $\Pi (q^2)$ are shown as a function of $q^2$.
As usual, there is a branch cut running form $q^2=4m_e ^2$ to the infinity.
Positronium is described as singular points below the mass threshold $4m_e ^2$,
and their positions are $q^2=M_N^2-\mathrm{i} M_N\Gamma _N$,
where $M_N$, $\Gamma _N$ are the mass and the decay width of the positronium state $N$, respectively.

There are infinite number of excited states of the positronium and
among them, spin-triplet $s$-wave states have the largest mixing with photon.
Considering the conservation of angular momentum, states with total angular momentum $j\not =1$ cannot mix with photon,
then the candidates for mixing are spin-singlet $p$--waves or spin-triplet $s$--waves.
In order to mix with the photon, electron and positron have to meet at zero separation, and
this means that the value of the wave function at the origin is related to the magnitude of mixing. 
The wave functions of $p$--waves are highly suppressed at the origin compared to those of $s$--waves,
so we concentrate on the mixing between photon and spin-triplet $s$-wave states of positronium in the following.

With the mixing with positronium explained above, the photon propagator acquires additional term
\begin{eqnarray}
\Pi (q^2) \sim \frac{Z_n q^2}{q^2 -M_n ^2 +\mathrm{i} M_n \Gamma _n} \label{PsPropagator}
\end{eqnarray}
where $Z_n$ is a mixing factor.
Note that we can label the relevant states by one index $n$, the principal quantum number.
Here and in the following,  $X\sim Y$ denotes that $X$ is equal to $Y$ near $q^2\simeq M_n^2$, i.e., except the regular parts.
The masses of the positronium are expressed as $M_n =2m_e -E_n$ where $E_n$ is the binding energy.
To the leading order of $\alpha$,
$E_n=\alpha ^2m_e /4n^2$.
The decay width of the $n=1$ state corresponds to the three-photons decay of ortho-positronium,
 $\Gamma _1=2(\pi^2-9)\alpha ^6m_e /9\pi$ \cite{Ore:1949te}.
For higher excited states, the decay width is not well known but here we assume $\Gamma_n\ll E_n$. 
In order to get an explicit form of $Z_n$, let us consider the right hand side of eq.\eqref{pi}
\begin{eqnarray}
&&\bra{0} {\rm T}\{j^\mu (x)j^\nu (0) \} \ket{0} \nonumber \\
&&= \theta (x^0)\bra{0} j^\mu (x)j^\nu (0)\ket{0} +\theta (-x^0)\bra{0} j^\nu (0)j^\mu (x)\ket{0}  \label{theta} 
\end{eqnarray}
where $\theta (x^0)$ is the step function.
Then we insert the complete set of the spin-triplet positronium states
\begin{eqnarray}
1\sim \sum _{n,\sigma }\int \frac{\mathrm{d} ^3 {\bf p} }{(2\pi )^3} \frac{1}{2E_{n,{\bf p} }}\ket{n,{\bf p},\sigma}\bra{n,{\bf p},\sigma} \label{1posi}
\end{eqnarray}
between two current operators.
In the eq.\eqref{1posi},  $E_{n,{\bf p}}$ is the energy of the positronium with mass $M_n$ and momentum ${\bf p}$, 
$\sigma$ is the polarization of the positronium, and the positronium one-particle state is
\begin{eqnarray}
\ket{n,{\bf p},\sigma}=\int \frac{\mathrm{d} ^3{\bf k}}{(2\pi )^3}
\sqrt{\frac{E_{n,{\bf p}}}{2E_1E_2}} ~\tilde \phi _n({\bf k} )\ket{{\bf k} _1 ,{\bf k}_2 ,\sigma} .
\end{eqnarray}
The right hand side is expressed by the states of electron positron pair,
with momentums ${\bf k} _1={\bf k} +{\bf p} /2$, ${\bf k} _2=-{\bf k} +{\bf p} /2$,
energies $E_1$, $E_2$,
and the spin configuration of the two particles is $\sigma $. 
The excitation of the positronium is described by $\tilde \phi _n ({\bf k} )$, the Coulomb wave function of $n$S state in the momentum space.
The insertion of eq.\eqref{1posi} into $\bra{0} j^\mu (x)j^\nu (0)\ket{0}$ leads to \footnote{
Similar calculation is done in ref.\cite{Peskin:1995ev} and ref.\cite{Ackleh:1991dy}.}
\begin{eqnarray}
&&\bra{0} j^\mu (x)j^\nu (0)\ket{0}\nonumber\\
&&\sim \sum _n 8\pi \alpha |\phi _n(0)|^2 \int \frac{\mathrm{d} ^3{\bf p} }{(2\pi )^3}  \left( g^{\mu\nu} -\frac{p^\mu p^\nu}{M_n^2}\right) \mathrm{e} ^{-\mathrm{i} px} \label{aa}
\end{eqnarray}
where $\phi _n(0)$ is the wave function at the origin in the coordinate space.
Note that  the positronium states are on-shell.
The factor $\left( g^{\mu\nu} -{p^\mu p^\nu}/{M_n^2}\right)$ comes from polarization sum of the positronium states,
and it ensures the gauge invariance at $p^2=M_n^2$.
Substituting eq.\eqref{aa} into eq.\eqref{theta}
and using the integral expression of the step function
\begin{eqnarray}
\theta (x^0)=-\int ^\infty _{-\infty}\frac{\mathrm{d} \omega}{(2\pi \mathrm{i} )} \frac{\mathrm{e} ^{-\mathrm{i} \omega x^0}}{\omega +\mathrm{i} 0^+} ,
\end{eqnarray}
we get the bound state contributions to the vacuum polarization function
\begin{eqnarray} \Pi (q^2) \sim \sum _n\frac{16\pi \alpha |\phi _n(0)|^2E_{n,{\bf p}}}{M_n^4} \frac{q^2}{q^2-M_n^2+\mathrm{i} 0^+} .\label{pspropagator}\end{eqnarray}
We use non-relativistic approximation to the positronium $E_{n,{\bf p}} \simeq M_n$ 
because the dominant contribution to the electron $g-2$ comes from non-relativistic momentum region.
Taking the width $\Gamma _n$ into account and comparing the eq.\eqref{PsPropagator} with eq.\eqref{pspropagator}, we finally obtain
\footnote{
The same result can be obtained by considering a scattering of hypothetical charged fermions, 
$X_1$ and $X_2$, that are lighter than the electron. 
The cross section near the positronium resonance, 
$\sigma(X_1 \bar{X}_1 \to {\rm Ps} \to X_2 \bar{X}_2)$, can be calculated in two ways; by using the Breit-Wigner formula, 
and by using eq.\eqref{PsPropagator}. By equating the two formulae, one obtains eq.\eqref{zn}.
}
\begin{eqnarray}
Z_n=\frac{16\pi \alpha |\phi _n(0)|^2}{M_n^3} \simeq \frac{\alpha ^4}{4n^3} .\label{zn}
\end{eqnarray}

Now, calculating the positronium contribution to $a_e=(g-2)_e/2$ is straightforward \cite{Jegerlehner:2009ry} and the result is
\begin{eqnarray}
\Delta a_e&=& \frac{\alpha }{\pi } \sum _n \left[ Z_n\ \int ^1 _0\mathrm{d} z \frac{z(1-z)^2m_e^2}{z M_n^2+(1-z)^2m_e^2} \right] \label{ae} \\
&=&\frac{\alpha ^5}{4\pi}\zeta (3) \times 0.045\\
& \simeq &9.0\times 10^{-14} .\label{result}
\end{eqnarray}

\section{Discussion}
In this section we show that the result of eq.\eqref{result} is distinct from perturbative contribution, 
and is not doubly counted in the finite order ladder diagrams.
Two ways to show that claim are given.

The first argument goes as follows.
Let us express the vacuum polarization contribution to $a_e$ as \cite{Jegerlehner:2009ry}
\begin{eqnarray}
\Delta a_e =-\frac{\alpha}{\pi ^2} \int ^\infty _{-\infty} \!\!\! \mathrm{d} s\ \frac{{\rm Im} \Pi (s)}{s} \int ^1 _0\!\! \frac{z(1-z)^2m_e^2\mathrm{d} z}{z s+(1-z)^2m_e^2} .
\end{eqnarray}
In the perturbative approach, the imaginary part of $\Pi (s)$ is nonzero only when $s\geq 4m_e ^2$ (FIG. \ref{plane}).
On the other hand, as we can see from eq.\eqref{PsPropagator} 
the imaginary part corresponding to the positronium comes from the singular point below $s=4m_e ^2$.
The distance between the singular point and the edge of the branch cut is about $4m_e E_n$, but
most of the contribution of the positronium comes from the range of $2m_e \Gamma _n$ around the pole and $\Gamma _n\ll E_n$.
Therefore the bound state contribution can be considered to be isolated from the perturbative one.

\begin{figure}[b!]
\begin{center}
\includegraphics[width =8cm, bb=72 641 540 720]{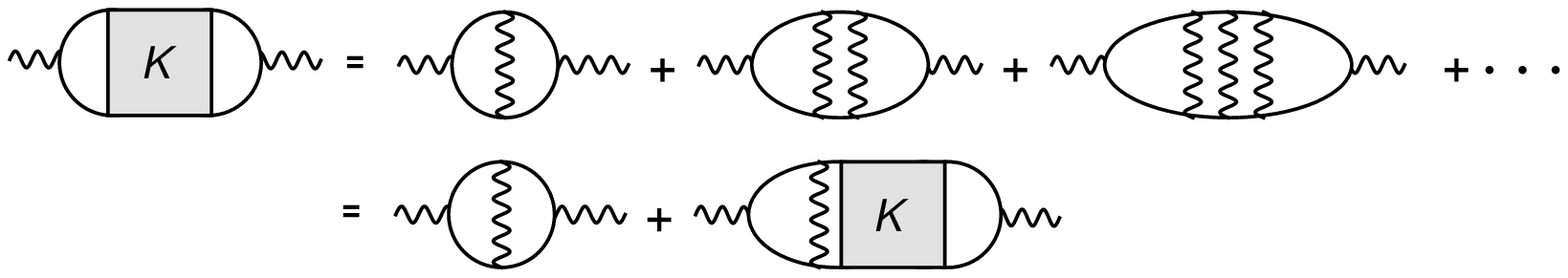}
\caption{Bethe-Salpeter equation beginning with single ladder diagram.
The gray box $K$ represents the two-particle Green function which includes ladder diagrams only.
The poles of $K$ below the mass threshold corespond to the bound states.
}
\label{bs1}
\includegraphics[width =8cm, bb=72 641 540 740]{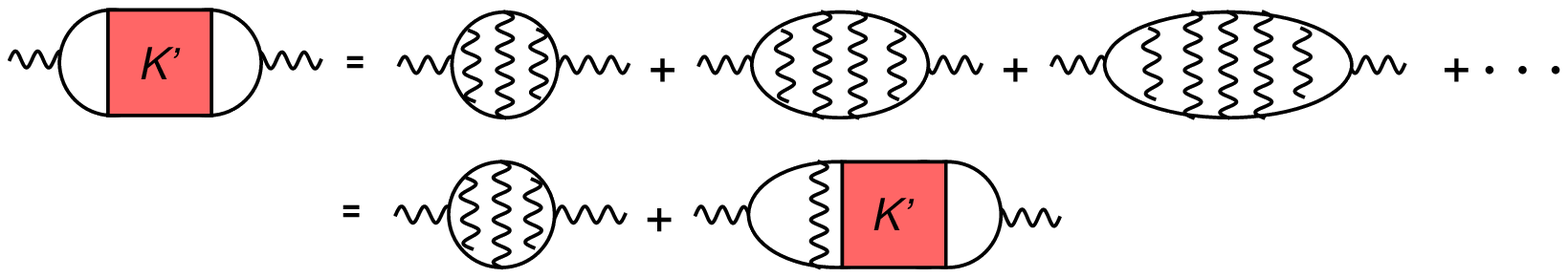}
\caption{Bethe-Salpeter equation which begins with triple ladder diagram.
The two-particle Green function $K'$ is different from $K$ of FIG. \ref{bs1},
but has same poles and residues as $K$.}
\label{bs2}
\end{center}
\end{figure}

The other way goes as follows.
What we have done in the previous section is equivalent to summing ladder diagrams using Bethe-Salpeter equation \cite{GellMann:1951rw}, \cite{Salpeter:1951sz}.
It is graphically expressed as FIG.~\ref{bs1}.
The important point is that the result of eq.\eqref{result} comes from the residue of $\Pi (q^2)$ at $q^2=M_n^2$,
and the residue is not influenced by diagrams of finite order.
For example, let us remove first two ladders from the upper equation of FIG.~\ref{bs1},
and construct Bethe-Salpeter equation which begins with triple ladder diagram (FIG.~\ref{bs2}).
Even in this case, the result is completely the same as eq.\eqref{result}.
In order to understand this fact with the simplest toy model,
let us consider an infinite sum with a perturbation parameter $a$
\begin{eqnarray}
f(x)&=&\frac{a}{x}+\frac{a^2}{x^2}+\frac{a^3}{x^3}+\cdots \nonumber\\
&=&\frac{a}{x-a} ,
\end{eqnarray}
which has a pole at $x=a$ and the residue is $a$.
Now, remove first ten terms, for instance, then the sum becomes
\begin{eqnarray}
f'(x)&=&\frac{a^{11}}{x^{11}}+\frac{a^{12}}{x^{12}}+\frac{a^{13}}{x^{13}}+\cdots \nonumber\\
&=&\left(\frac{a}{x}\right)^{10}\frac{a}{x-a} .
\end{eqnarray}
We can find that the residue at $x=a$ is unchanged.
In short, if we concatenate only on the singular part, the first few orders of the infinite sum is irrelevant.
Poles and residues are generated from the operation of ``infinite sum" itself.
This is why the contribution from positronium is distinct from the contribution from the perturbation of ``finite" order.

Our contribution \eqref{result} is of order $\alpha ^5$, but it came from a non-perturbative effect.
Perturbation theory produces terms of the form $\alpha ^n$, but not all terms of the form $\alpha ^n$ are produced by perturbation theory alone.

Finally, let us estimate the uncertainty of our result.
What we discussed in this latter is the leading contribution of the virtual positronium,
and there are radiative corrections to it.
However those contributions are suppressed by an extra factor $(\alpha /\pi )$ and supposed to make little effect.
If we consider $(\alpha /\pi )^6$ order, there is another contribution which is topologically different from the vacuum polarization correction.
Other source of uncertainty is relativistic correction to $M_n$ and $\phi _n(x)$ but these are all suppressed by $\alpha$.
Also we have to comment on the assumption $\Gamma _n\ll E_n$ used above.
This condition is satisfied for small $n$ but we cannot assure that this holds for all excited states of the positronium.
Fortunately, the higher excited states contribute not so much to $a_e$ because they are suppressed by the factor $n^{-3}$.
All these taken into account, we set conservative estimate of 0.1 as the relative uncertainty.

\section{Conclusion}
We show that there is a new contribution to the electron $g-2$, which originates from the virtual positronium propagation.
This contribution can be distinguished from ordinary perturbative one.
In the TABLE 1, we summarize the standard model prediction of $a_e$ and compare it with the measured value.
The non-perturbative QED contribution we estimated in this letter is about one order of magnitude smaller than the current theoretical uncertainty
but larger than the electroweak contribution.
Works are in progress for a reduction of the uncertainties of both theory and experiment \cite{Giudice:2012ms},
so this virtual positronium effect may become visible in the future.

\renewcommand{\arraystretch}{1.2}
\begin{table*}
\begin{ruledtabular}
\begin{tabular}{l r r c}
&value $\ (\times 10^{14}$)&uncertainty$\ (\times 10^{14}$)&references\\\hline
perturbative QED&115 965 218 007$\!\! \quad$&77$\!\! \quad$&\cite{Aoyama:2012wj}\\
non-perturbative QED&9.0&0.9&this work\\
hadronic&167.8&1.4&\cite{Nomura:2012sb}, \cite{Prades:2009tw}\\
electroweak&3.0&0.1&\cite{Mohr:2012tt}\\\hline
standard model prediction&115 965 218 187$\!\! \quad$&77$\!\! \quad$&\\\hline
measurement&115 965 218 073$\!\! \quad$&28$\!\! \quad$&\cite{2008PhRvL.100l0801H}\\\hline
difference&114$\!\! \quad$&82$\!\! \quad$&\\
\end{tabular}
\caption{Summary of the standard model prediction of $a_e$. 
The fifth line is the total of first four contributions.
For comparison, we list the measured value and the difference between them.
The uncertainty of the difference is the geometric mean of the theoretical uncertainty and the measured uncertainty.
}
\end{ruledtabular}
\end{table*}

\section{Acknowledgments}
The author is very grateful to Koichi~Hamaguchi, Yuji~Tachikawa, Motoi~Endo, and Kyohei~Mukaida
for fruitful discussions, instructive suggestions, and hearty encouragement.

\appendix
\section{Appendix}
The bound state effect on the muon $g-2$ is calculated in exactly the same way as that of electron,
and the result is
\begin{eqnarray} \Delta a_\mu = 9.6\times 10^{-13}, \label{muon} \end{eqnarray}
which is about two orders of magnitude smaller than the current theoretical and experimental uncertainty.

\bibliography{mishima.bib}

\end{document}